\documentclass[aps,reprint,amsmath,amssymb,superscriptaddress,nofootinbib,floatfix]{revtex4-2}
\usepackage{graphicx}
\usepackage{booktabs}
\usepackage{mathrsfs}
\usepackage[colorlinks=true,allcolors=blue]{hyperref}
\graphicspath{{figures/}}
\newcommand{\Msun}{M_\odot}
\begin{document}

\title{Rotational tidal susceptibility of white dwarfs at fixed baryon number}
\author{Edson Otoniel}
\email{edson.otoniel@ufca.edu.br}
\affiliation{Universidade Federal do Cariri, Instituto de Forma\c{c}\~ao de Educadores, Brejo Santo, CE, Brazil}
\author{Jonathan A. Rebou\c{c}as}
\email{jalvesreboucas@ifce.edu.br}
\affiliation{Instituto Federal de Educa\c{c}\~ao, Ci\^encia e Tecnologia do Cear\'a, Iguatu, CE, Brazil}
\author{Fridolin Weber}
\email{fweber@sdsu.edu}
\affiliation{Department of Physics, San Diego State University,
San Diego, CA 92182, USA}
\affiliation{Department of Physics, University of California,
San Diego, La Jolla, CA 92093, USA}
\date{\today}

\begin{abstract}
We calculate the stationary aligned quadrupolar tidal response of cold
carbon fluid white dwarfs in general relativity, retaining uniform
rotation through second order. Equilibrium sequences conserve physical
baryon number, while the tidal perturbations also conserve angular
momentum. The dimensionless susceptibility to rotation near zero spin
increases from $0.844$ to $5.443$ between initial masses of $0.6$ and
$1.3M_\odot$. Independent partial derivatives show that this dependence
is mainly structural: a larger central pressure adjustment combines
with a steeper dependence of the spherical deformability on pressure.
The explicit mixed correction reduces the response at $0.6$ and
$1.0M_\odot$ but increases it at $1.3M_\odot$. Although the total
deformability grows in every admitted model, its local zero spin
approximation increasingly underestimates the finite spin enhancement.
Density maps and material-surface embeddings connect this response to
the equilibrium readjustment without identifying a stellar outline
with a tidal deformability.
\end{abstract}
\maketitle

\section{Introduction}\label{sec:introduction}

White dwarfs connect the physics of degenerate matter to stellar
structure and evolution~\cite{Chandrasekhar1931,Althaus2010}. Their
occurrence in compact binaries also motivates calculations of the
response to an external gravitational field~\cite{Korol2017,Kupfer2024}.
Tidal deformability depends on both the size of a star and its internal
response. The fifth power of the radius in the spherical coefficient
makes this distinction especially relevant when rotation is varied.
A model at fixed central density and one containing the same amount of
matter as an initial static star answer different questions. The latter
requires the central pressure to adjust, so a change in the tidal
coefficient combines a change of background structure with an explicit
coupling between rotation and the applied tide.

Relativistic equilibria through second order in angular velocity are
described by the Hartle expansion~\cite{Hartle1967,HartleThorne1968}.
Applications to white dwarfs include sequences of constant rest mass
and fixed composition~\cite{Boshkayev2013}. Studies of rapidly spinning
white dwarfs also show why rotational support must be accounted for
when interpreting a mass and radius~\cite{Otoniel2021Rotation}.
Rotational changes in the mass--radius relation can also reduce the
separation between ordinary white dwarfs and strange
dwarfs~\cite{Otoniel2026StrangeRotation}.
Conservation of matter and the associated structural readjustment are
therefore established ingredients of rotating stellar modelling.
White dwarf equilibrium has separately been examined with matter-trace
couplings~\cite{Carvalho2017}, matter-Lagrangian
couplings~\cite{Otoniel2026FRTL}, and quadratic
torsion~\cite{Otoniel2026Torsion}. These studies motivate a clearly
specified general relativistic baseline: changes in a dimensional
response must first be distinguished from those produced by rotation
and the choice of equilibrium sequence within the same gravity theory.

Relations between moment of inertia, tidal Love number and rotational
quadrupole were developed for neutron and quark
stars~\cite{YagiYunes2013}, and subsequently investigated in white
dwarfs~\cite{Boshkayev2017,Taylor2020}. These quantities describe
different properties: the isolated quadrupole created by rotation is
not the quadrupole induced by an external tide. Relativistic tidal
theory further distinguishes multipolar response coefficients from
surface-shape coefficients~\cite{BinningtonPoisson2009,DamourNagar2009}.
For white dwarfs, composition and elastic stresses provide additional
sources of variation; calculations of crystallized stars show that a
fluid response cannot be applied indiscriminately to an elastic
core~\cite{PerotChamel2022}. Fixing the material EOS and the fluid
assumption isolates the rotational dependence considered here.

Spin also enters tidal perturbations directly. Relativistic exterior
calculations identify couplings between rotation and the applied
field that require additional response
coefficients~\cite{Pani2015,LandryPoisson2015Exterior}.
Stationary axisymmetric tidal coefficients have been calculated for
neutron stars at linear order in spin~\cite{Pani2015SpinLove}, while
interior calculations require a specified fluid
state~\cite{Landry2017}. A stationary external field alone does not
exclude internal motion in every tidal sector, as illustrated by the
response to a gravitomagnetic field~\cite{LandryPoisson2015Dynamical}.
Here we retain the stationary aligned electric quadrupole and allow a
uniformly rotating fluid to adjust at fixed baryon number and angular
momentum during each applied tidal perturbation. This specifies a
restricted response convention, rather than the full tide of an
orbiting companion.

Dynamical tides involve different physics. Excitation and
damping of gravity waves enter white dwarf synchronization and
dissipation~\cite{FullerLai2012}; resonance locks couple spin and
orbital evolution~\cite{Burkart2013}. Rotation changes the dynamical
tidal response~\cite{FullerLai2014}, and nonlinear mode interactions
can modify the subsequent evolution~\cite{YuWeinbergFuller2020}.
Normal-mode formulations for rotating fluids provide the corresponding
frequency dependent description~\cite{Schenk2002,Pnigouras2024}.
A stationary deformability supplies neither these transfer rates nor
an oscillation spectrum. This distinction allows an equilibrium
calculation to identify structural effects without attributing them to
a resonance or a dissipative mechanism.

Building on the rotational tidal formulation for white dwarfs in
$f(R,T)$ gravity~\cite{Otoniel2026SlowRotation}, we examine how the
initial mass affects the response at fixed baryon number in general
relativity. A single cold carbon equation of state (EOS) provides a
controlled comparison of three reference masses, $0.6$, $1.0$ and
$1.3M_\odot$. We calculate the
local rotational susceptibility from independent static and rotating
partial derivatives, separate the required background readjustment
from the explicit mixed tidal term, and compare that local coefficient
with finite rotating equilibria. This comparison quantifies the mass
dependence of the two contributions and tests where a local coefficient
adequately represents the calculated sequences. Density reconstructions
and surface embeddings give complementary spatial interpretations.
The paper develops the model and its boundary conditions in
Sec.~\ref{sec:framework}, presents the response and geometry
results in Sec.~\ref{sec:results}, and discusses their implications and
extensions in Sec.~\ref{sec:conclusions}.

\section{Theoretical framework}\label{sec:framework}

\subsection{Carbon matter and spherical equilibria}

We use general relativity, metric signature $(-,+,+,+)$ and $G=c=1$.
Pressure $p$ and energy density $\epsilon$ refer to the physical fluid,
with $T_{ab}=(\epsilon+p)u_a u_b+pg_{ab}$. The composition is fixed to
fully ionized $^{12}\mathrm C$ at zero temperature. Charge neutrality
relates the electron, ion and baryon densities by $n_e=6n_i=n_B/2$.
Retaining $\hbar$ explicitly, electron degeneracy gives
\begin{equation}
\begin{aligned}
n_e&=\frac{p_F^3}{3\pi^2\hbar^3},\qquad E(q)=\sqrt{m_e^2+q^2},\\
\epsilon_{\rm el}&=\frac{1}{\pi^2\hbar^3}\int_0^{p_F}E(q)q^2\,dq,\\
p_{\rm el}&=\frac{1}{3\pi^2\hbar^3}\int_0^{p_F}\frac{q^4}{E(q)}\,dq.
\end{aligned}\label{eq:electrons}
\end{equation}
Here $p_F$ is the Fermi momentum and $m_e$ the electron rest mass.
Ionic rest energy supplies most of $\epsilon$, while electrons supply
the degeneracy pressure. The electrostatic lattice contribution obeys
$\epsilon_L\propto-n_i^{4/3}$ and $p_L=\epsilon_L/3$~\cite{Salpeter1961}.
Thus $p=p_{\rm el}+p_L$ and
$\epsilon=\epsilon_{\rm ion}+\epsilon_{\rm el}+\epsilon_L$, with
electron rest energy counted only once. The lattice contributes to the
isotropic EOS; elastic shear stresses, magnetic
fields and composition changes are absent. Strong magnetic fields can
introduce pressure anisotropy through Landau
quantization~\cite{Otoniel2026MagneticAnisotropy}; these effects lie
outside the isotropic fluid model adopted here. Cold stellar models with
Coulomb corrections~\cite{HamadaSalpeter1961} and more detailed
microscopic treatments~\cite{Rotondo2011} establish the sensitivity to
the material prescription. Changes in electron kinematics can modify
white dwarf structure while leaving the general relativistic field
equations unchanged~\cite{Otoniel2026Dispersion}.
We keep the carbon EOS specified above fixed throughout the comparison.
Extending a fixed composition to higher density
would also require checking electron-capture
thresholds~\cite{ChamelFantina2015Capture}.
Density values in $\mathrm{g\,cm^{-3}}$ denote
$\rho=\epsilon_{\rm phys}/c^2$, with $c$ restored for this conversion.

For the spherical background, write
$ds_0^2=-e^\nu dt^2+w^{-1}dr^2+r^2d\Omega_2^2$, where
$w=1-2m/r$. The structure equations are~\cite{Tolman1939,OppenheimerVolkoff1939}
\begin{equation}
\begin{aligned}
m'&=4\pi r^2\epsilon,\\
\frac{\nu'}2\equiv g_0&=\frac{m+4\pi r^3p}{r^2w},
\qquad p'=-(\epsilon+p)g_0.
\end{aligned}\label{eq:tov}
\end{equation}
Radial derivatives are denoted by primes until pressure derivatives are
introduced explicitly below. Regular central series initialize the
integration, with $m(0)=0$ and $p(0)=p_c$; they are not an expansion of
the entire stellar solution. At the material endpoint,
$M_0=m(R_0)$ and $e^{\nu(R_0)}=1-2M_0/R_0$. The common surface
prescription uses $p_{\rm cut}=3.27571803\times10^{-21}
\,\mathrm{MeV\,fm^{-3}}$. Its negligible residual pressure is omitted
in vacuum matching, with $4\pi R_0^3p_{\rm cut}/M_0$ monitored in
geometrized units. Finite surface density terms are retained.
No atmosphere or material shell is added.

\subsection{Uniform rotation and conserved matter}

Uniform circular flow has $u^a=u^t(1,0,0,\Omega)$. In the Hartle gauge,
the metric through second order in rotation is~\cite{Hartle1967}
\begin{equation}
\begin{aligned}
ds^2={}&-e^\nu(1+2h)dt^2+
 \frac{1+2\mathfrak m/(rw)}{w}\,dr^2\\
&+r^2(1+2k)\{d\theta^2+\sin^2\theta(d\phi-\omega dt)^2\}.
\end{aligned}\label{eq:metric}
\end{equation}
For the isolated star, with $P_\ell$ the Legendre polynomials,
$h=h_0+h_2P_2$, $\mathfrak m=m_0+m_2P_2$,
$k=k_2^{\rm met}P_2$ and $\omega=O(\Omega)$; the other perturbations
are $O(\Omega^2)$. The angular metric coefficient $k_2^{\rm met}$ is
distinct from the tidal Love number. With $j=e^{-\nu/2}\sqrt w$ and
$\bar\omega=\Omega-\omega$, dragging satisfies
\begin{equation}
(r^4j\bar\omega')'+4r^3j'\bar\omega=0.\label{eq:dragging}
\end{equation}
Central regularity gives $\bar\omega'(0)=0$. Matching to
$\bar\omega=\Omega-2J/r^3$ fixes
$J=R_0^4\bar\omega'(R_0)/6$ and
$\Omega=\bar\omega(R_0)+2J/R_0^3$.

Define $p_\ell^*=\delta p_\ell/(\epsilon+p)$, with $\delta p_\ell$
the Eulerian pressure perturbation, and
\begin{equation}
T_a=\frac{r^4j^2(\bar\omega')^2}{6},\quad
T_b=-\frac{r^3(j^2)'\bar\omega^2}{3},\quad T=T_a+T_b.
\label{eq:spin-sources}
\end{equation}
The monopole equations are
\begin{equation}
\begin{aligned}
m_0'={}&4\pi r^2\frac{\epsilon+p}{c_s^2}p_0^*+\frac{T_a}{2}+T_b,\\
(p_0^*)'={}&-\frac{1+8\pi r^2p}{r^2w^2}m_0
 -\frac{4\pi r(\epsilon+p)}{w}p_0^*\\
&+\frac{T_a}{2rw}
 +\frac13\left(\frac{r^2j^2\bar\omega^2}{w}\right)',
\end{aligned}\label{eq:spin-monopole}
\end{equation}
where $c_s^2=dp/d\epsilon$ and $p_0^*(0)=0$ at prescribed $p_c$.
Writing $v_2=h_2+k_2^{\rm met}$ and
\begin{equation}
\begin{aligned}
\mathcal A&=-\nu'+\frac{8\pi(\epsilon+p)-4m/r^3}{w\nu'},\\
\mathcal B_\ell&=-\frac{\ell(\ell+1)-2}{r^2w\nu'},
\end{aligned}\label{eq:polar-coefficients}
\end{equation}
the quadrupole sector takes the compact form
\begin{equation}
\begin{aligned}
h_2'&=\mathcal A h_2+\mathcal B_2v_2
       +\frac{\nu'T}{2}+\frac{T_b-T_a}{r^2w\nu'},\\
v_2'&=-\nu'h_2+\left(\frac1r+\frac{\nu'}2\right)T,\\
m_2&=rw(-h_2+T).
\end{aligned}\label{eq:spin-quadrupole}
\end{equation}
Regular particular and homogeneous solutions are matched to the isolated
vacuum exterior. The Euler integral gives
$p_2^*=-h_2-r^2e^{-\nu}\bar\omega^2/3$ and the isobar displacements
$\xi_\ell=p_\ell^*/g_0$. Labelling isobars by their radius $a$ in
the associated spherical background, the interior radial chart is
$r(a,\mu)=a+\xi_0(a)+\xi_2(a)P_2(\mu)$, with $\mu=\cos\theta$.
The full mass correction includes the displaced surface~\cite{ReinaVera2015}:
\begin{equation}
M_{\rm rot}=M_0+m_0(R_0)+\frac{J^2}{R_0^3}
                +4\pi R_0^2\epsilon(R_0)\xi_0(R_0).
\label{eq:mass-correction}
\end{equation}

On a constant time slice bounded by the displaced material surface
$r=R_\Sigma(\mu)$, the physical baryon count and angular momentum are
evaluated as
\begin{equation}
\begin{aligned}
A&=2\pi\int_{-1}^{1}d\mu\int_0^{R_\Sigma(\mu)}
                  n_Bu^t\sqrt{-g}\,dr,\\
J&=2\pi\int_{-1}^{1}d\mu\int_0^{R_\Sigma(\mu)}
          (\epsilon+p)u^tu_\phi\sqrt{-g}\,dr.
\end{aligned}\label{eq:charges}
\end{equation}
The determinant here is in $(t,r,\mu,\phi)$ coordinates.
At zero spin, $A_0=4\pi\int_0^{R_0}n_Br^2w^{-1/2}dr$.
Its rotational correction at prescribed central pressure is
\begin{equation}
\begin{aligned}
\Omega^2A_2={}&4\pi\int_0^{R_0}\frac{r^2}{\sqrt w}
 \left[\delta n_{B0}+n_B\left(\frac{m_0}{rw}
             +\frac{r^2e^{-\nu}\bar\omega^2}{3}\right)\right]dr\\
&+\frac{4\pi R_0^2n_B(R_0)\xi_0(R_0)}{\sqrt{w(R_0)}},\\
\delta n_{B0}={}&\frac{dn_B}{dp}(\epsilon+p)p_0^*.
\end{aligned}\label{eq:baryon-correction}
\end{equation}
The subscript in $A_2$ denotes rotational order, not multipolarity.
Each isolated sequence solves $A(p_c,\Omega)=A_{\rm ref}$ for $p_c$,
where $A_{\rm ref}$ is the initial static baryon count; $J$ varies along
that sequence. Conservation of $A$ and $J$ during an applied tidal
perturbation is a separate constraint.

\subsection{Stationary quadrupolar tidal response}

We use the general relativistic limit of the stationary fixed-$A,J$
formulation of Ref.~\cite{Otoniel2026SlowRotation}.
Choose the spin axis as $z$ and an applied gravitational tidal tensor
$\mathcal E_{ij}=\mathrm{diag}(-\mathcal E/2,-\mathcal E/2,\mathcal E)$.
Its potential is $U_{\rm ext}=\mathcal E r^2P_2(\mu)/2$, so that the
leading applied metric term is $\delta g_{tt}=-\mathcal E r^2P_2$.
We retain the stationary aligned component $\ell=2,m=0$. It defines a
restricted response coefficient, not every component of a rotating
star's tidal tensor. For the spherical reference,
$Q_{ij}^{\rm ind}=-\lambda_0\mathcal E_{ij}$ and
$\lambda_0=(2/3)k_2R_0^5$~\cite{FlanaganHinderer2008}.

With $H(r)$ the radial amplitude of the static quadrupolar metric
perturbation and $y=rH'/H$, the spherical quadrupole equation is
\begin{equation}
\begin{aligned}
ry'+y^2+Fy+r^2\mathcal V&=0,\\
F&=\frac{1+4\pi r^2(p-\epsilon)}{w},\\
r^2\mathcal V&=\frac{4\pi r^2[5\epsilon+9p+(\epsilon+p)/c_s^2]-6}{w}\\
&\quad-4\left(\frac{m+4\pi r^3p}{rw}\right)^2.
\end{aligned}\label{eq:static-tide}
\end{equation}
The regular solution approaches $y=2$ at the centre. The exterior
growing and decaying solutions are proportional to
$P_2^2(r/M_0-1)$ and $Q_2^2(r/M_0-1)$, the associated Legendre functions
of the first and second kinds, both of degree and order two~\cite{Hinderer2008}.
The finite surface density enters through
$y_R=y(R_0^-)-4\pi R_0^3\epsilon(R_0)/M_0$ under the adopted vacuum
prescription~\cite{TakatsyKovacs2020}. For $\mathscr C=M_0/R_0$ and
$B=2-y_R+2\mathscr C(y_R-1)$, their matching gives
\begin{equation}
\begin{aligned}
k_2&=\frac{8\mathscr C^5(1-2\mathscr C)^2B}{5\mathcal D_k},\\
\mathcal D_k&=2\mathscr C[6-3y_R+3\mathscr C(5y_R-8)]\\
&\quad+4\mathscr C^3[13-11y_R+\mathscr C(3y_R-2)
                    +2\mathscr C^2(1+y_R)]\\
&\quad+3(1-2\mathscr C)^2B\ln(1-2\mathscr C),
\end{aligned}\label{eq:love}
\end{equation}
using the corrected static expression~\cite{Hinderer2009Erratum}.

Rotation couples this polar quadrupole to axial $\ell=1,3$ fields at
$O(\mathcal E\Omega)$ and polar $\ell=0,2,4$ fields at
$O(\mathcal E\Omega^2)$~\cite{Pani2015}. To specify the sources, let
$s$ and $\tau$ count powers of rotation and tide. In Eq.~\eqref{eq:metric},
\begin{equation}
\begin{aligned}
h={}&s^2(h_{{\rm s}0}+h_{{\rm s}2}P_2)+\tau h_{{\rm t}2}P_2
       +s^2\tau\sum_{\ell=0,2,4}h_{{\rm st}\ell}P_\ell,\\
\omega={}&s\omega_{\rm s}
       +s\tau\sum_{\ell=1,3}\omega_{{\rm t}\ell}P_\ell'(\mu),\\
\Omega={}&s\Omega_{\rm s}+s\tau\delta\Omega_1.
\end{aligned}\label{eq:hierarchy}
\end{equation}
In the isolated sequences, $\Omega=\Omega_{\rm s}$ ($s=1,\tau=0$);
$\delta\Omega_1$ is the tidal change in fluid angular velocity at first
rotational order.
The function $\mathfrak m$ has the same polar structure as $h$;
$k$ has polar $\ell=2$ at each lower order and $\ell=2,4$ at mixed
order, with no monopole. Primes on the Legendre polynomials denote
derivatives with respect to $\mu$. All products and inverse metrics
are expanded through $s^2\tau$, with no quadratic tide.

Fluid coefficients follow from the uniform-flow first integral
$W-\ln u^t=C_E$, where $dW=dp/(\epsilon+p)$ and
$u^t=[-(g_{tt}+2\Omega g_{t\phi}+\Omega^2g_{\phi\phi})]^{-1/2}$.
For any $\mathcal X(p)$, including $p$, $\epsilon$ and $n_B$, its mixed
coefficient is
\begin{equation}
\mathcal X_{\rm st}=\mathcal X_W W_{\rm st}+\mathcal X_{WW}W_{\rm s}W_{\rm t},\quad
\mathcal X_W=(\epsilon+p)\mathcal X_p,
\label{eq:material}
\end{equation}
where $\mathcal X_{WW}=(\epsilon+p)^2\mathcal X_{pp}
+(\epsilon+p)(1+c_s^{-2})\mathcal X_p$. Subscripts s, t and st identify
$s^2$, $\tau$ and $s^2\tau$ coefficients of the enthalpy, Eulerian
pressure and radial displacement perturbations;
$\xi_{\rm s}=\xi_0+\xi_2P_2$. This fixes the matter sources without a
separate density ansatz.

We express the radial equations using an explicit residual prescription.
Set the unknown fields at the order being solved to zero and extract
the corresponding known coefficient of
$\mathcal R_{ab}=G_{ab}-8\pi T_{ab}$ from
Eqs.~\eqref{eq:metric}, \eqref{eq:hierarchy} and \eqref{eq:material}.
For $\Pi_\ell[f]=(2\ell+1)\int_{-1}^{1}P_\ell f\,d\mu/2$, define
$\mathscr R_\ell=\Pi_\ell[\mathcal R_{rr}]$,
$\mathscr T_0=\Pi_0[\mathcal R^t{}_t]$ and
\begin{equation}
\begin{aligned}
\mathscr V_\ell&=\frac{2\ell+1}{2\ell(\ell+1)}
  \int_{-1}^{1}(1-\mu^2)P_\ell'\mathcal R_{r\mu}\,d\mu,\\
\mathscr U_\ell&=\frac{(2\ell+1)(\ell-2)!}{2(\ell+2)!}
  \int_{-1}^{1}(1-\mu^2)P_\ell''
  \left(\frac{\mathcal R_{\mu\mu}}{g_{\mu\mu}}
       -\frac{\mathcal R_{\phi\phi}}{g_{\phi\phi}}\right)d\mu.
\end{aligned}\label{eq:projections}
\end{equation}
Here the polar sources are extracted at $s^2\tau$. For $\ell=2,4$,
the inhomogeneous systems are
\begin{equation}
\begin{aligned}
h_{{\rm st}\ell}'&=\mathcal A h_{{\rm st}\ell}
                  +\mathcal B_\ell v_{{\rm st}\ell}+f_{h\ell},\\
v_{{\rm st}\ell}'&=-\nu'h_{{\rm st}\ell}+f_{v\ell},\qquad
m_{{\rm st}\ell}=-rw h_{{\rm st}\ell}+f_{m\ell},\\
f_{h\ell}&=\frac{\nu'r^2}{2}\mathscr U_\ell
  +\left(1+\frac{2}{r\nu'}\right)\mathscr V_\ell
  +\frac{\mathscr R_\ell}{\nu'},\\
f_{v\ell}&=(r+\nu'r^2/2)\mathscr U_\ell+\mathscr V_\ell,\qquad
f_{m\ell}=r^3w\mathscr U_\ell,
\end{aligned}\label{eq:mixed-polar}
\end{equation}
where $v_{{\rm st}\ell}=h_{{\rm st}\ell}+k_{{\rm st}\ell}$.
At order $s\tau$, expand
$\mathcal R^t{}_{\phi}=(1-\mu^2)\sum_{\ell=1,3}q_\ell P_\ell'$.
The axial equations are
\begin{equation}
\begin{aligned}
\omega_{{\rm t}\ell}''={}&
 \left[\frac{\ell(\ell+1)-2}{r^2w}+\frac{16\pi(\epsilon+p)}{w}\right]
       \omega_{{\rm t}\ell}\\
&+\left[-\frac4r+\frac{4\pi r(\epsilon+p)}{w}\right]
       \omega_{{\rm t}\ell}'
 \\
&-\delta_{\ell1}\frac{16\pi(\epsilon+p)}{w}\delta\Omega_1
 +\frac{2e^\nu q_\ell}{r^2w},
\end{aligned}\label{eq:mixed-axial}
\end{equation}
with $\ell=1,3$ and $\delta_{\ell1}$ the Kronecker symbol.
Their solved fields enter the $s^2\tau$ residual, not just the vacuum
matching. For the mixed monopole, define
$a_0=4\pi r(\epsilon+p)/w$,
$b_0=(1+8\pi r^2p)/(r^2w^2)$ and
$d_0=4\pi r^2(\epsilon+p)/c_s^2$. Then
\begin{equation}
\begin{aligned}
h_{{\rm st}0}'&=-a_0h_{{\rm st}0}+b_0m_{{\rm st}0}
                  +a_0\delta C_2-\frac r2\mathscr R_0,\\
m_{{\rm st}0}'&=-d_0h_{{\rm st}0}+d_0\delta C_2
                  +\frac{r^2}{2}\mathscr T_0.
\end{aligned}\label{eq:mixed-monopole}
\end{equation}
The constants $\delta\Omega_1$ and $\delta C_2$ (the $s^2\tau$
coefficient of $C_E$) are fixed by
$\delta J=0$ and $\delta A=0$, including the moving integration domain
in Eq.~\eqref{eq:charges}. Regular homogeneous polar fields scale as
$h_\ell\propto r^\ell$, $v_\ell\propto r^{\ell+2}$ and axial fields
as $\omega_\ell\propto r^{\ell-1}$. Particular solutions retain the
lower powers forced by the preceding orders.

At the material boundary, the Lagrangian pressure perturbation obeys
$\Delta p=0$. Its mixed coefficient includes
$\delta p_{\rm st}+p'\xi_{\rm st}
+\xi_{\rm s}\delta p_{\rm t}'+\xi_{\rm t}\delta p_{\rm s}'
+p''\xi_{\rm s}\xi_{\rm t}=0$ at $R_0$. Induced metric and extrinsic
curvature are matched on this displaced surface, not at a fixed
coordinate radius. The complete canonical applied mode is retained in
that matching and subtracted only when extracting the asymptotically
flat response~\cite{Pani2015}. The invariant mass quadrupole $M_2$ is
obtained from the circular-geodesic expansion~\cite{Ryan1995}, with
$Q_{zz}=2M_2/3$ in the trace-free Cartesian convention. Removing the
isolated rotational moment defines
\begin{equation}
\lambda_T=-\left.\frac{\partial Q_{zz}^{\rm ind}}
 {\partial\mathcal E}\right|_{A,J}
 =\lambda_0(p_c)+\delta\lambda_{\rm mix}(p_c,\Omega).
\label{eq:tidal-response}
\end{equation}
It has units of length to the fifth power. Holding $A,J$ fixed selects
the adjustment of a uniformly rotating fluid; it is not a dynamical
response with independently prescribed circulation.

For Fig.~\ref{fig:surface}, the angular metric on a constant time material
surface has
$\gamma_{\theta\theta}=[g_{\theta\theta}+g_{rr}(dR_\Sigma/d\theta)^2]_
{r=R_\Sigma}$ and $\gamma_{\phi\phi}=[g_{\phi\phi}]_{r=R_\Sigma}$.
Its Euclidean embedding satisfies
\begin{equation}
X=\sqrt{\gamma_{\phi\phi}},\qquad
\frac{dZ_{\rm emb}}{d\theta}
 =-\sqrt{\gamma_{\theta\theta}-(dX/d\theta)^2},
\label{eq:embedding}
\end{equation}
with $Z_{\rm emb}(\pi/2)=0$ and reflection completing the meridian.
Products and square roots retain the same perturbative orders as the
stellar fields. The prescribed forcing is
$\eta=\mathcal E R_0^3/M_0$; $X$ and $Z_{\rm emb}$ use the fixed
initial $R_{\rm ref}$ only when plotted.

\subsection{Local susceptibility and numerical construction}

Let $p_{c0}$ identify the initial static star, with
$M_{\rm ref}=M_0(p_{c0})$, $R_{\rm ref}=R_0(p_{c0})$ and
$\lambda_{\rm ref}=\lambda_0(p_{c0})$, with $A_{\rm ref}=A_0(p_{c0})$.
At prescribed central
pressure, expand the isolated baryon count and the tidal response as
\begin{equation}
\begin{aligned}
A(p_c,\Omega)&=A_0(p_c)+\Omega^2A_2(p_c)+O(\Omega^4),\\
\lambda_T(p_c,\Omega)&=\lambda_0(p_c)+\Omega^2L_2(p_c)+O(\Omega^4).
\end{aligned}\label{eq:partials}
\end{equation}
The subscript in $L_2$ likewise denotes rotational order. This
coefficient includes the constraints on the applied tidal derivative
in Eq.~\eqref{eq:tidal-response}. If $A_0'\ne0$,
enforcing $A=A_0(p_{c0})$ gives
$p_c=p_{c0}-\Omega^2A_2/A_0'+O(\Omega^4)$. Here and in the following
formula, primes on $A_0$ and $\lambda_0$ are derivatives with respect
to $p_c$ at $p_{c0}$:
\begin{equation}
\begin{aligned}
D_A&=\left.\frac{\partial\lambda_T}{\partial\Omega^2}\right|_{A,0}
       =L_2-\frac{A_2}{A_0'}\lambda_0',\\
\chi_A&=\frac{M_{\rm ref}}{R_{\rm ref}^3\lambda_{\rm ref}}D_A,
\qquad \epsilon_{\rm ref}=\frac{\Omega^2R_{\rm ref}^3}{M_{\rm ref}}.
\end{aligned}\label{eq:chi}
\end{equation}
Consequently $\lambda_T/\lambda_{\rm ref}-1
=\chi_A\epsilon_{\rm ref}+O(\Omega^4)$ locally.
Here $[A_2]=L^2$, $[L_2]=[D_A]=L^7$ and $\chi_A$ is dimensionless.
The structural and mixed terms in Eq.~\eqref{eq:chi} define
$\chi_{\rm str}$ and $\chi_{\rm mix}$. At finite spin we use
\begin{equation}
S=\frac{\lambda_0[p_c(\Omega)]}{\lambda_{\rm ref}}-1,
\quad C=\frac{\delta\lambda_{\rm mix}}{\lambda_{\rm ref}},
\quad \frac{\lambda_T}{\lambda_{\rm ref}}-1=S+C.
\label{eq:decomposition}
\end{equation}
Differentiating the structural term separates the radius and Love number
contributions through $d\ln\lambda_0=d\ln k_2+5d\ln R_0$.

We determine $A_2$ directly from the physical current, without subtracting
nearly equal total baryon counts, and obtain $L_2$ from two independent
rotating calculations at the initial static central pressure. They use
the physical angular velocities of the $x=0.05$ and $0.10$ sequence
members, where $x=\Omega/\Omega_K$ and $\Omega_K$ is the Kepler
reference frequency associated with each equilibrium, varying along the sequence.
The auxiliary calculations need not have those same $x$ labels.
Static derivatives use three symmetric steps
$\Delta\ln p_c=0.002$, $0.001$ and $0.0005$.
At finite spin, we solve the retained baryon integral for
$p_c$ at each spin, without expanding that root about $p_{c0}$.
This choice preserves the finite readjustment while retaining the same
second order field equations.

The spherical structure equations determine the matter and metric
profiles entering the perturbation hierarchy. Regular homogeneous and
particular solutions are integrated from the centre and matched to the
vacuum exterior on the displaced material surface. Matching and
conservation of baryon number and angular momentum fix the tidal
amplitudes and fluid readjustment. Axial tidal fields linear in rotation
enter the mixed polar sources at second order.
Logarithmic interpolation of the
fixed EOS gives $\epsilon=K_ip^{\alpha_i}$ and
$n_B=N_ip^{\beta_i}$ within each interval. Pressure derivatives in the
sources are taken from this same interpolant. Derivative jumps at its
knots produce interface contributions in the mixed equations and
physical baryon current; they are integrated explicitly and are not
physical density transitions. Finite surface contributions are treated
separately. Exterior matching is evaluated at arbitrary precision because
the small compactness of white dwarfs causes cancellations in the
response basis.

Each point has its own radial/EOS, central-start, angular, baryon-pairing
and low/high surface-cut controls. Their absolute changes are combined
as the sum of the first four sensitivities plus the larger surface-cut
shift, separately for each observable. Physical-current quadratures,
static derivative steps, the independent probe discrepancy and reference
normalizations enter the susceptibility envelope. These are empirical
numerical envelopes, not statistical intervals or bounds on all omitted
terms. We require total refinement changes below $10^{-4}$, changes
relative to the static reference below $10^{-6}$, and relative mixed
refinement changes below $10^{-3}$. The restrictions
$|\delta A|/A<10^{-5}$, $|\delta J|/J<10^{-5}$ and surface residuals
below $10^{-10}$ are checked, together with field linearity and spin
parity at $10^{-8}$. The largest isolated baryon pairing residual is
$1.7\times10^{-8}$. We call an effect resolved when its magnitude
exceeds three times its own envelope. The spherical problem is
cross-checked using metric variables and the Riccati equation, with
agreement within $10^{-5}$.

We obtain $\Omega_K$ by equating the stellar angular velocity to the
prograde circular-geodesic frequency at the equatorial surface of the
second order metric. It is not used
to assert accuracy near mass shedding. Our tidal sample has
$\Omega^2R_0^3/M_0\leq0.1$ and metric/displacement amplitudes below
$0.1$; this background screening differs from
$\epsilon_{\rm ref}$. Spins $x=0,0.05,0.10,0.20,0.30,0.35$ are
admitted for all three masses, while $x=0.40$ is admitted only for
$1.0$ and $1.3\Msun$. Neither this screening nor numerical refinement
certifies the size of the omitted physical rotational orders.

\section{Results and discussion}\label{sec:results}

\begin{figure}[tb]
\includegraphics[width=\columnwidth]{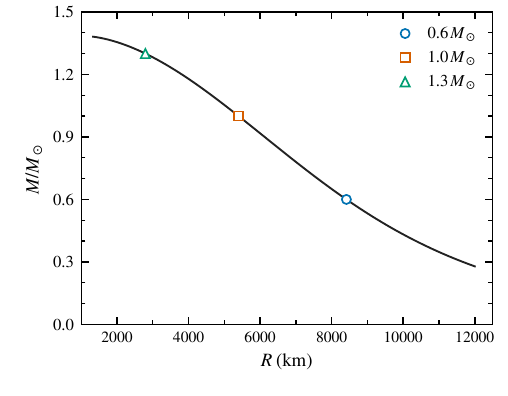}
\caption{Static mass--radius sequence for the adopted cold carbon fluid.
The symbols select the initial models of $0.6$, $1.0$ and
$1.3\Msun$, whose baryon numbers are conserved along the respective
rotating sequences. The black curve connects directly integrated static
equilibria within the adopted density domain and $R\leq12000\,\mathrm{km}$;
its endpoints are sampling limits, not a stability classification.}
\label{fig:selection}
\end{figure}

The initial configurations span the static mass--radius sequence shown in
Fig.~\ref{fig:selection}. Their reference masses are
$M_{\rm ref}=0.6$, $1.0$ and $1.3\Msun$, with radii of approximately
$8412$, $5397$ and $2791\,\mathrm{km}$, respectively. Each reference
defines its own conserved baryon number; the three rotating sequences
therefore compare different amounts of matter within the same carbon
equation of state. Rotation changes the gravitational mass, central
pressure and spherical background associated with each equilibrium.
Neither the mass label nor the normalization radius is updated along a
sequence. This distinction permits a comparison of the response to
rotational support without mixing it with a change of composition or an
implicit reselection of the reference star.

\begin{figure*}[tb]
\includegraphics[width=\textwidth]{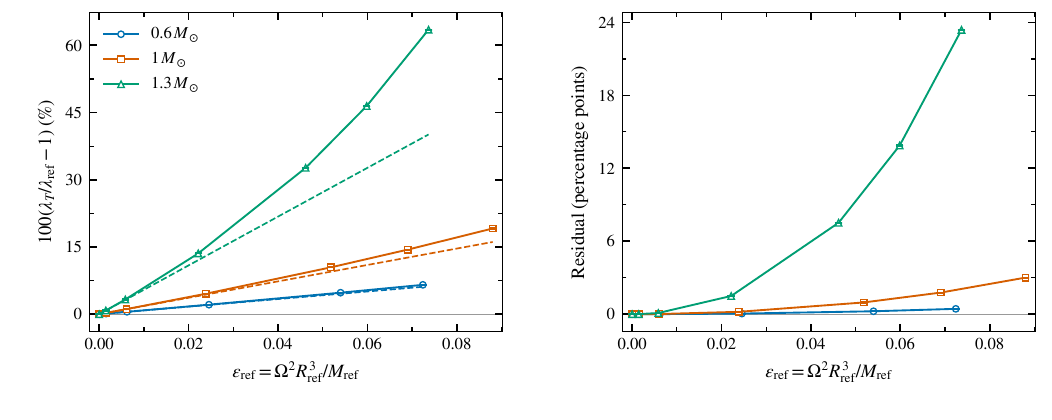}
\caption{Response along sequences of fixed physical baryon number.
Left: fractional change in the dimensional deformability relative to
the initial static value. Symbols indicate calculated equilibria and
solid segments connect these nodes. Dashed lines give the independent
local prediction $100\chi_A\epsilon_{\rm ref}$, with its propagated
numerical envelope. Right: direct response minus this prediction, in
percentage points. Bars show empirical numerical envelopes and may be
smaller than the symbols. The abscissa uses the fixed initial mass and
radius, $\epsilon_{\rm ref}=\Omega^2R_{\rm ref}^3/M_{\rm ref}$, with
$G=c=1$. The residual is not an estimate of omitted physical terms of
fourth order in rotation.}
\label{fig:response}
\end{figure*}

At fixed baryon number, the dimensional tidal deformability increases at
every calculated positive spin in all three sequences
(Fig.~\ref{fig:response}). The increase depends strongly on the
reference mass. At $x=0.35$, the enhancements relative to
the respective static values are $6.55\%$, $14.41\%$ and $46.46\%$.
The corresponding values of $\epsilon_{\rm ref}$ are $0.0724$, $0.0690$
and $0.0599$. Thus the larger fractional response of the most massive
sequence is not explained by a larger value of the adopted rotational
strength. The same ordering is already present in the local susceptibility
$\chi_A$, obtained independently from the static derivatives and the
rotating partial derivatives at fixed central pressure.
Table~\ref{tab:susceptibility} reports its values; the coefficient for
$1.3\Msun$ is about $6.45$ times that for $0.6\Msun$.

\begin{table}[tb]
\caption{Local rotational susceptibility and its decomposition at the
three reference masses. The quoted $U_A$ is a propagated empirical
numerical envelope, not a statistical uncertainty. Both independent
small spin probes agree within their own envelopes. Component signs are resolved when
$|\chi_{\rm str}|>3U_{\rm str}$ and $|\chi_{\rm mix}|>3U_{\rm mix}$,
using their respective envelopes. All coefficients are dimensionless.}
\label{tab:susceptibility}
\begin{ruledtabular}
\begin{tabular}{cccc}
$M_{\rm ref}/\Msun$ & $\chi_{\rm str}$ & $\chi_{\rm mix}$ & $\chi_A\pm U_A$\\
\hline
0.6 & 0.9602 & $-0.1166$ & $0.8436\pm0.0002$\\
1.0 & 1.8431 & $-0.0130$ & $1.8301\pm0.0003$\\
1.3 & 5.3199 & $+0.1229$ & $5.4428\pm0.0024$
\end{tabular}
\end{ruledtabular}
\end{table}

Agreement between the two probes used to calculate $\chi_A$ does not
make the tangent at zero spin an accurate description of an entire
sequence. The right panel of Fig.~\ref{fig:response} shows the departure
from this tangent. At $x=0.35$, the local prediction underestimates the
enhancement by $0.44$, $1.78$ and $13.88$ percentage points, in order of
increasing reference mass. These discrepancies exceed the propagated
numerical envelopes. The admitted extensions to $x=0.40$ give total
enhancements of $19.12\%$ and $63.50\%$ for the two heavier sequences;
the $0.6\Msun$ configuration at that spin falls outside the declared
rotational domain and is omitted. The finite spin calculations solve
the retained baryon constraint for the adjusted central pressure at each
spin, while the local prediction expands that adjustment about the
initial static state. Their discrepancy measures the importance of this
finite readjustment within the adopted construction. It does not
determine the missing fourth order contribution to the physical stellar
solution, which would require an extension of the rotational equations.

To identify the origin of the mass dependence, consider baryon
conservation before the explicit mixed tidal term. Central pressure
must decrease because rotation increases the amount of matter supported
at a prescribed central pressure. In the notation of
Eq.~\eqref{eq:chi}, this is the contribution
$-(A_2/A_0')\lambda_0'$ to $D_A$. Both $A_0'$ and $A_2$ are positive
for the reference configurations, whereas $\lambda_0'$ is negative.
The pressure adjustment therefore increases the spherical
deformability. Quantitatively, the derivative
$d\ln p_c/d\epsilon_{\rm ref}$ at zero spin changes from approximately
$-1.40$ to $-4.89$ between $0.6$ and $1.3\Msun$. Over the same mass
interval, $d\ln\lambda_0/d\ln p_c$ changes from $-0.688$ to $-1.087$.
A larger pressure readjustment and a steeper dependence of the spherical
response on pressure therefore act together. Their product explains the
increase in $\chi_{\rm str}$ in Table~\ref{tab:susceptibility}; the
denominator $A_0'$ by itself is not a stability criterion.

\begin{figure*}[tb]
\includegraphics[width=\textwidth]{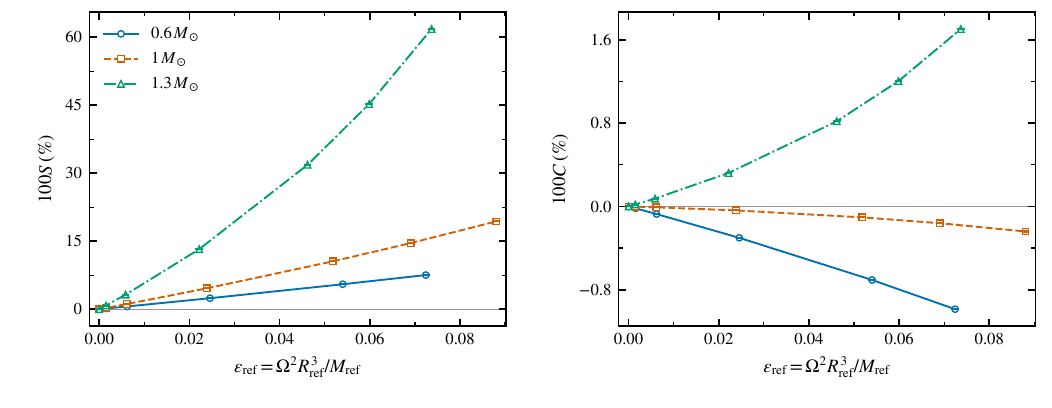}
\caption{Structural and explicit mixed contributions to the response
in Fig.~\ref{fig:response}. Left: $100S$, where
$S=\lambda_0[p_c(\Omega)]/\lambda_{\rm ref}-1$. Right:
$100C=100\delta\lambda_{\rm mix}/\lambda_{\rm ref}$.
The total fractional change is $S+C$. Colours identify the same
reference masses in both panels; symbols are computed equilibria and
lines connect the nodes. Bars represent the individual numerical
envelopes. The vertical scales differ, so a comparable visual excursion
does not imply comparable contributions.}
\label{fig:decomposition}
\end{figure*}

Figure~\ref{fig:decomposition} extends this separation to finite spin.
The structural term remains positive and supplies most of the total
change. Its radius and Love number contributions can be distinguished
through $\lambda_0=(2/3)k_2R_0^5$, using the radius of the associated
spherical background. In the local coefficient, the radius contribution
accounts for $89.3\%$, $85.7\%$ and $85.6\%$ of
$\chi_{\rm str}$ for the three masses. Changes in $k_2$ supply the
remaining positive contribution. An estimate based only on the fifth
power of a radius would therefore miss part of the structural response.
Replacing $R_0$ with the equatorial radius would introduce a different
prescription, since the latter already contains the angular deformation
of the rotating equilibrium.

An explicit coupling between rotation and the applied tide remains after
this background adjustment. Its sign differs between the sampled masses:
the mixed term reduces the deformability for $0.6$ and $1.0\Msun$, and
increases it for $1.3\Msun$. This sign difference appears in both the
local coefficients and the finite spin curves, and is resolved by their
numerical envelopes. For the lightest sequence, the negative mixed term
reduces the total enhancement by about $15.1\%$ at $x=0.35$; the
structural term is correspondingly larger than the total. For the
heaviest sequence, the mixed term contributes about $2.6\%$ of the
total enhancement at the same fractional Kepler spin. A positive total
response thus does not require a positive explicit mixed correction.
Three reference masses do not locate a continuous zero of that
correction or establish its behaviour at other compositions. The split
also depends on the specified spherical background and applied-field
convention; its two terms are not separately measurable tidal
observables.

\begin{figure*}[tb]
\includegraphics[width=\textwidth]{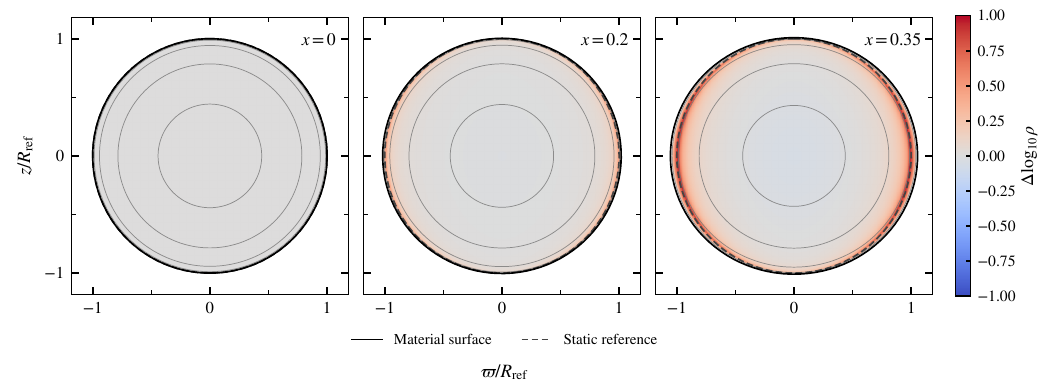}
\caption{Density redistribution along the sequence starting from
$1.0\Msun$, at $x=0$, $0.20$ and $0.35$. Both axes use the fixed
initial radius $R_{\rm ref}=5397.4\,\mathrm{km}$ and the interior
Hartle radial chart, with $z=r\cos\theta$ and $|\varpi|=r\sin\theta$;
the section is mirrored to negative $\varpi$. Colour gives
$\Delta\log_{10}\rho=\log_{10}[\max(\rho_\Omega,\rho_f)/\max(\rho_0,\rho_f)]$
at the same coordinate position, with rotating and initial static profiles
$\rho_\Omega$ and $\rho_0$, and plotting floor
$\rho_f=10^4\,\mathrm{g\,cm^{-3}}$.
Thin grey contours indicate
$\log_{10}[\rho/(\mathrm{g\,cm^{-3}})]=4,5,6,7$.
Solid and dashed curves mark the current and initial static material
surfaces. The maps compare stationary equilibria, not a time dependent
transport of matter.}
\label{fig:density}
\end{figure*}

Density maps in Fig.~\ref{fig:density} show the spatial changes
associated with the central pressure adjustment along the representative
$1.0\Msun$ sequence at fixed baryon number. Its central mass density
decreases from $3.412\times10^7$ to
$3.056\times10^7\,\mathrm{g\,cm^{-3}}$ between zero spin and
$x=0.35$. At that spin, the equatorial and polar radii in the interior
radial chart are approximately $5690.7$ and $5445.0\,\mathrm{km}$,
both larger than the initial $5397.4\,\mathrm{km}$. The configuration
remains oblate because its equatorial radius exceeds its polar radius;
oblateness does not require polar contraction relative to the initial
star. Positive colour near the outer equatorial region compares matter
in the expanded star with the lower density at the same coordinate
position in the reference. It is not a calculation of mass flow into a
crust. The reconstruction carries the EOS density on displaced isobars,
with no additional envelope, and retains the isotropic fluid model used
in the response calculation.

\begin{figure*}[tb]
\includegraphics[width=\textwidth]{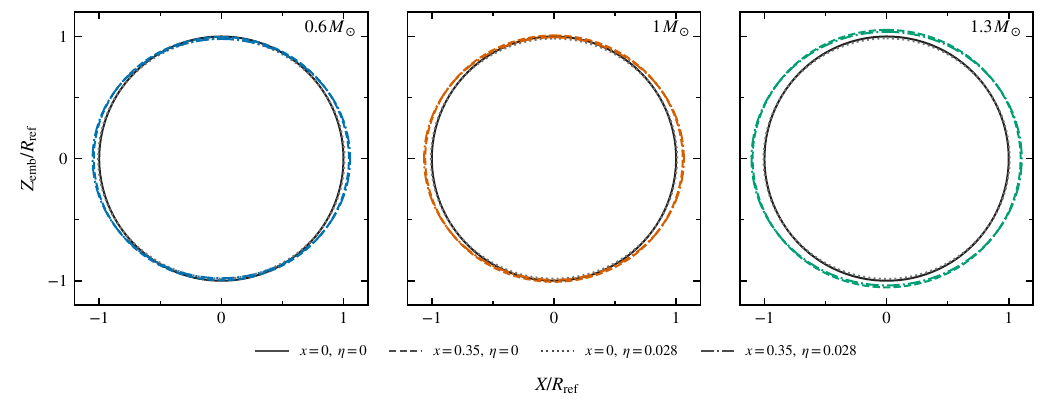}
\caption{Meridional embeddings of the material-surface angular metric
for the three reference masses. Styles distinguish the four combinations
of $x=0,0.35$ and prescribed $\eta=0,0.028$.
Coloured rotating contours use the mass colours of
Fig.~\ref{fig:response}. Axes are normalized by each initial static
radius. The forcing is $\eta=\mathcal E R_0^3/M_0$, using the
associated spherical background, so equal $\eta$ does not impose equal
dimensional fields. Its prescribed value follows the scale discussed in
the text. The embedding includes the material displacement and angular
metric through $\Omega^2$, linear order in $\mathcal E$, and their
mixed order. No distortion amplification is applied; $Z_{\rm emb}$ is
an embedding height, not a coordinate polar radius.}
\label{fig:surface}
\end{figure*}

A surface under a specified external field is a different diagnostic
from either the density map or the deformability coefficient.
Figure~\ref{fig:surface} reconstructs the intrinsic angular metric on
the displaced material surface, rather than identifying the induced
quadrupole with an outline. The prescribed value $\eta=0.028$ is rounded
from $1/(3.3)^3$, using the equal mass theoretical configuration with
periastron separation $D\simeq3.3R$ and component radius $R$ in Fig.~1 of
McNeill et al.~\cite{McNeillMardlingMueller2020}. The conversion
$\eta=(M_{\rm comp}/M_0)(R_0/D)^3$, with $M_{\rm comp}$ the companion
mass, uses the leading Newtonian $m=0$ projection of the tidal potential~\cite{YuWeinbergFuller2020}.
We adopt only this forcing scale; no orbit or companion is inferred
for the isolated models, and the stationary aligned field omits the
other azimuthal components of a binary tide.

At $x=0.35$, this field increases the equatorial circumferential radius
relative to the isolated rotating configuration by approximately
$0.85$--$0.90\%$ across the three masses, and decreases the embedded
polar height by approximately $1.46$--$1.64\%$. Large changes in the
coefficient $\lambda_T$ can therefore coexist with a small additional
surface distortion at a specified forcing. The comparison also depends
on normalization: all contours use their own fixed initial
$R_{\rm ref}$, while the same dimensionless $\eta$ corresponds to
different dimensional fields for different masses and backgrounds.
Neither similarity of normalized contours nor the dominance of a radius
term in the susceptibility makes the tidal response a purely geometric
function of the stellar outline. Its calculation still requires the
fluid perturbations, the conservation constraints and the exterior
multipole matching.

\section{Conclusions}\label{sec:conclusions}

Conserving the amount of matter changes the interpretation of rotational
tidal response. For cold carbon fluid white dwarfs in general relativity,
the local susceptibility increases from $0.8436$ to $5.4428$ between
initial masses of $0.6$ and $1.3\Msun$. These coefficients compare
sequences with distinct, individually conserved baryon numbers; they do
not describe stars held at the same central density. The response
therefore depends on both the initial stellar mass and the physical
constraint used to identify successive rotating equilibria.

Independent partial derivatives identify the main origin of this mass
dependence. The more massive reference requires a larger decrease of
central pressure per unit rotational strength, while its spherical
deformability has a steeper pressure dependence. Radius expansion
supplies most of the resulting structural contribution, with a positive
change of the spherical Love number supplying the remainder. Once this
readjustment is separated, the explicit mixed correction reduces the
response in the two lighter sequences and increases it in the heaviest
one. The signs are resolved by the numerical controls, but three masses
do not locate the transition continuously or establish a universal
relation.

Across all admitted rotating models, the total deformability increases.
Its independently calculated tangent at zero spin increasingly
underestimates the response at finite spin, especially for the most
massive sequence. Agreement between two small spin probes thus does
not establish the usefulness of a local coefficient throughout an
entire sequence. Here the discrepancy concerns the implicit equilibrium
readjustment within the retained expansion; it is not a measurement of
the omitted fourth order physics. Comparison with models retaining the
full rotational deformation is needed to determine whether these finite
spin trends persist beyond the present approximation.

Density maps and material-surface embeddings connect the response to
the calculated equilibria. The maps describe differences between
stationary density profiles, without implying time dependent transport
of matter. Under the prescribed field $\eta=0.028$, the embeddings show
that a large fractional change in deformability can coexist with a
small additional surface distortion. Neither a coordinate contour nor
the isolated rotational quadrupole determines the tidal coefficient.
The material displacement, angular metric, fluid adjustment and
exterior matching remain distinct ingredients of the response.

Quantitative conclusions apply to the admitted domain of uniform
rotation through second order and a stationary aligned quadrupolar tide
in an isotropic carbon fluid. Repeating the derivative calculation at
additional baryon targets would test the mass dependence and locate
the change of sign of the mixed coefficient. Composition and elastic
stresses would require a corresponding extension of the matter model,
rather than a reinterpretation of the fluid curves. A frequency
dependent treatment must additionally specify the internal motion and
its coupling to orbital evolution before addressing resonances,
dissipation or binary signals. Those calculations would assess whether
the structural and explicit mixed contributions identified here remain
distinguishable when the star is driven dynamically.

\begin{acknowledgments}
Edson Otoniel acknowledges support from Funda\c{c}\~ao Cearense de Apoio ao Desenvolvimento Cient\'ifico e Tecnol\'ogico (FUNCAP) through grant BP6-0241-00335.01.00/25. 
\end{acknowledgments}

\bibliographystyle{apsrev4-2}
\bibliography{ref}
\end{document}